\documentclass[prd, amsmath, floats, floatfix, twocolumn, nopreprintnumbers, superscriptaddress, nofootinbib, eqsecnum]{revtex4}

\usepackage{ulem}
\usepackage{color} 
\usepackage{graphicx} 
\usepackage{float} 
\usepackage{amsmath} 
\usepackage{acronym} 
\usepackage{multirow} 
\usepackage{verbatim} 
\usepackage{subfigure}
\usepackage[squaren]{SIunits}

\addunit{\Mparsec}{Mpc}
\addunit{\smass}{M_{\odot}}
\addunit{\yr}{yr}

\newcommand{\bea}{\begin{eqnarray}} 
\newcommand{\eea}{\end{eqnarray}}

\newcommand{\beq}{\begin{equation}} 
\newcommand{\eeq}{\end{equation}}

\def\ltsima{$\; \buildrel < \over \sim \;$}
\def\simlt{\lower.5ex\hbox{\ltsima}}
\def\gtsima{$\; \buildrel > \over \sim \;$}
\def\simgt{\lower.5ex\hbox{\gtsima}}

\begin{document}

\title{Studies of waveform requirements for intermediate mass-ratio coalescence searches 
with advanced gravitational-wave detectors}

\author{R. J. E. Smith} 
\affiliation{School of Physics and Astronomy, University of Birmingham, Edgbaston, 
Birmingham B15 2TT, UK} 
\affiliation{Perimeter Institute for Theoretical Physics, Waterloo, Ontario N2L 2Y5, Canada}

\author{I. Mandel} 
\affiliation{School of Physics and Astronomy, University of Birmingham, Edgbaston, 
Birmingham B15 2TT, UK}

\author{A. Vecchio} 
\affiliation{School of Physics and Astronomy, University of Birmingham, Edgbaston, 
Birmingham B15 2TT, UK}

\begin{abstract}

The coalescence of a stellar-mass compact object into an intermediate-mass black hole 
(intermediate mass-ratio coalescence; IMRAC) is an important astrophysical source for 
ground-based gravitational-wave interferometers in the so-called advanced (or second-generation) configuration.
However, the ability to carry out effective matched-filter based searches for these systems is limited by the lack of reliable waveforms. Here we consider binaries in which the intermediate-mass black hole has mass in the range $24\, \smass - 200\, \smass$ with a stellar-mass companion having masses in the range $1.4\, \smass - 18.5\, \smass$. In addition, we constrain the mass ratios, $q$, of the binaries to be in the range $1/140 \leq q \leq 1/10$ and we restrict our study to the case of circular binaries with non-spinning components. We investigate the relative contribution to the signal-to-noise ratio (SNR) of the three different phases of the coalescence: inspiral, merger and ringdown using waveforms computed within the effective one-body formalism matched to numerical relativity, known as EOBNR. We show that merger and ringdown contribute to a substantial fraction of the total SNR over a large portion of the mass parameter space, although in a limited portion the SNR is dominated by the inspiral phase. We further identify three regions in the IMRAC mass-space in which: (i) inspiral-only searches could be performed with losses in detection rates $L$ in the range $10\% \lesssim L \lesssim 27\%$, (ii) searches based on inspiral-only templates lead to a loss in detection rates in the range $27\% \lesssim L \lesssim 50\%$, and (iii) templates that include merger and ringdown are essential to prevent losses in detection rates greater than $50\%$. In addition we find that using inspiral-only templates as filters can lead to large biases in the estimates of the mass parameters of IMRACs. We investigate the effectiveness with which the inspiral-only portion of the IMRAC waveform space is covered by comparing several existing waveform families in this regime. We find that different waveform families are only marginally effective at describing one another, as measured by the ``fitting factor''. Our results reinforce the importance of extensive numerical relativity simulations of IMRACs to validate and calibrate semi-analytical waveform families and the need for further studies of suitable approximation schemes in this mass range.
\end{abstract}

\maketitle 
\section{ Introduction}

Observations of ultra-luminous X-ray sources and simulations of globular cluster dynamics 
suggest the existence of intermediate-mass black holes (IMBHs). However, observational evidence for their existence is still under debate, see e.g.~\cite{MillerLISA:2009,MillerIMBH:2004}. Gravitational waves from binary coalescences involving IMBHs with masses 
\unit{50}{\smass} $\lesssim M \lesssim$ \unit{500}{\smass} are potentially detectable by advanced 
detectors -- including Advanced LIGO \cite{Harry:2010}, Advanced Virgo \cite{aVirgo}, and KAGRA \cite{KAGRA}  -- with a low frequency cutoff of around \unit{10}{\hertz}. 
If IMBHs do exist, one likely contribution to gravitational-wave detections %rates of binary coalescences 
is believed to be through the coalescence of a compact stellar-mass companion 
(black hole or neutron star) with an IMBH, at a possible rate of up to $\sim$\unit{10}{\yr^{-1}} \cite{LSCVrates:2010,BrownIMRI:2007,MandelIMRI:2008}. 
We will denote these signals as intermediate mass-ratio coalescences (IMRACs)\footnote{
In the literature, the term frequently used for this class of objects is \textit{intermediate
mass-ratio inspirals} or IMRIs, see e.g.~\cite{BrownIMRI:2007,MandelIMRI:2008}. However, in the context of
ground-based observations, in particular with second-generation instruments, we will
show that the full coalescence is important for these systems, and it therefore seems to be more
appropriate to call them IMRAC.}.
%These are known as intermediate mass-ratio coalescences (IMRACs). 
%Simulations of various 
%formation mechanisms suggest that three-body hardening is likely the dominant mode of 
%formation. Under this mechanism IMRACs are likely to have circularized by the time they 
%enter into the frequency band of ground based detectors. This results in eccentricities 
%$e\lesssim10^{-5}$ \cite{MandelIMRI:2008}, ensuring the small body follows quasi- 
%circular orbits which endows the emitted gravitational waves with the classic ``chirp'' 
%signature. 

Given that IMBHs in this mass range have proved extremely difficult 
to observe in the electromagnetic spectrum, gravitational-wave detections may provide the first unambiguous observations of such objects through the robust measurement of their masses. Such observations would form an important channel for probing the dynamical history of globular clusters. Furthermore, Advanced LIGO/Virgo (aLIGO/Virgo) may be 
able to provide measurements of the quadrupole moment of a black hole
\cite{BrownIMRI:2007, Rodriguez:2012}, which would allow a null-hypothesis test of the Kerr metric for IMBHs.

The gravitational waveform from the coalescence of two compact objects can be divided into three phases: a gradual inspiral, a rapid merger, and the quasi-normal ringdown of the resulting black hole. The relative contribution to the expected coalescence signal from inspiral, merger and ringdown is an important consideration for gravitational-wave searches.
To leading Newtonian order the gravitational wave frequency at the inner-most stable circular orbit (ISCO) is 
\beq 
f_{\mathrm{ISCO}} = \unit{4.4}k{\hertz}\ \left(\frac{M_{\odot}}{M}\right)\,, 
\label{eq:fisco} 
\eeq 
where $M$ is the total mass of the binary. For advanced detectors with a low frequency 
cut-off of $\sim 10\,\mathrm{Hz}$, we may only have access to either the very late stages 
of the inspiral, or solely merger and ringdown for the heaviest IMRAC systems. While the 
power in the merger and ringdown is suppressed by a factor of the mass ratio relative to the power in the inspiral, the fact that IMRAC systems are liable to merge either in-band, or at 
the low frequency limit of the bandwidth, means that merger and ringdown may be 
significant over a large portion of the detectable mass-space. Additionally, for cases where 
IMRAC waveforms are inspiral-dominated, it is useful to know where inspiral-only searches 
could be targeted. 

Detecting IMRACs through gravitational waves will require template gravitational-waveform families adapted to highly asymmetrical mass-ratio systems.
However the development of numerical relativity simulations and perturbative techniques in this regime is at an early stage which is potentially problematic (see, e.g. \cite{NRpert} for a discussion of this issue).  
The issue of appropriate template waveform families is thus central to the detection of IMRACs through gravitational waves. 

The effective-one-body approach, calibrated to numerical relativity, has led to template waveforms, known as EOBNR 
\cite{PanEOBNR:2011}, that describe the full inspiral, merger and ringdown coalescence-signal for 
comparable mass-ratio binaries; EOBNR waveforms should also be accurate at extreme mass ratios. However, to date only one full numerical simulation exists 
for mass-ratio $q=1/100$ binaries \cite{RIT:2011}. Furthermore, EOBNR waveforms have been constructed to reproduce the dynamical evolution of binaries with extreme mass ratios. EOBNR waveforms have not yet been compared to numerical relativity simulations at such mass ratios, so their validity in the IMRAC regime remains to be demonstrated. 

Meanwhile, in the context of extreme mass-ratio binaries, several authors have modelled the two-body motion by computing 
radiative and conservative self-force corrections to Kerr geodesic motion 
\cite{GG:2006, BFGGH:2007}. Waveforms computed 
within this scheme are inspiral-only and are only developed to lowest order in the mass ratio. 
These waveforms have been adapted to describe intermediate mass-ratio inspirals by including higher-order-in-mass-ratio corrections in \cite{HuertaGair:2009} and have been used to
study the detection of intermediate mass-ratio inspirals in the context of the proposed third-generation ground-based gravitational-wave interferometer the Einstein Telescope \cite{HG:2011}.  We refer to 
these intermediate mass-ratio inspiral waveforms as the ``Huerta Gair'' (HG) waveform family after its authors. 
This waveform family should be physically well motivated to describe the inspiral of IMRACs.

Typically one does not have an exact representation of ``true'' gravitational-wave signals but requires templates which are sufficiently
effective at filtering such signals.  A  common metric for quantifying how well
approximate waveform families are at filtering gravitational-wave signals is known the ``effectiveness'', or fitting factor \cite{Buonanno:2009}. 
This measures the fraction of the theoretical maximum signal-to-noise ratio (SNR) that could be recovered by using non-exact template waveforms.

The work in this paper proceeds as follows. Firstly, by computing the effectiveness of inspiral-only template waveforms at filtering the full coalescence signal, 
we determine the relative importance of the inspiral and merger-ringdown phases. We identify three regions in the component mass plane in which: 
$(a)$ inspiral-only searches are feasible with losses in detection rates $L$ in the range $10\% \lesssim L \lesssim 27\%$, $(b)$ searches are limited
by the lack of merger and ringdown in template waveforms and are liable to incur losses in detection rates in the range $27\% \lesssim L \lesssim 50\%$, and $(c)$ merger and ringdown are essential for searches in order to prevent losses in detection rates greater than $50\%$. We also study the biases incurred in estimates of the mass parameters of IMRACs when using inspiral-only waveforms to filter signals which are described by full inspiral, merger and ringdown waveforms. We find that estimates in the chirp mass and symmetric mass ratio of IMRACs can be biased by as much as $50\%$ and $160\%$, respectively, of the values encoded in the signal waveform.

Secondly, to gain insight into the accuracy of the inspiral portion of IMRAC waveforms we compute the effectiveness of the inspiral-only 
portion of EOBNR waveforms at filtering gravitational-wave signals as described by the HG waveform family. We find that there is a non-negligible 
discrepancy between EOBNR and HG inspirals in the regime where inspiral-only searches could be considered sufficient. For reference we also compare EOBNR 
inspirals to a post-Newtonian (PN) waveform family known as TaylorT4 \cite{Buonanno:2009}. The PN expansion is liable to be a poor choice of approximant 
for IMRACs because of the large number of cycles spent at small radii. We find that EOBNR 
and HG are in better agreement with each other than to TaylorT4, as might be expected from the previous 
observation. 

Our approach does not directly address the accuracy of template 
waveforms, because none of the waveforms considered have been matched to full 
numerical waveforms.  However, assuming that the waveform families we consider ``bracket'' the correct gravitational waveforms in the IMRAC regime, this approach provides
 a useful heuristic for quantifying the effectiveness of existing gravitational waveforms for IMRAC searches.
Further numerical relativity simulations will be important in the continuing development of accurate template waveforms for IMRACs. 

Our analysis improves upon previous work to determine the detectability of 
IMRAC sources \cite{MandelGair:2009} which only considered the so-called ``faithfulness'' of 
template waveforms, i.e., the effectiveness of  template waveforms evaluated at the signal
parameters. Additionally, that study only considered inspiral-only waveforms and focused on low frequency observations, e.g. with the proposed Laser Interferometer Space Antenna (LISA).

This paper is organized as follows. In Sec.~\ref{sec:waveforms} we describe the waveform families used in our study. In Sec.~\ref{sec:SNR} we compute the contributions to the SNR from the 
inspiral and merger and ringdown phases of EOBNR waveforms in the intermediate mass-ratio
regime. In Sec.~\ref{sec:IMR} we study the effectiveness of inspiral-only waveforms to filter full coalescence signals from 
IMRAC sources and identify the three regions in which different searches could be conducted. In Sec.~\ref{sec:param_bias} we study biases incurred in estimates of the mass parameters of IMRAC sources through using inspiral-only templates to filter full inspiral, merger and ring-down waveforms.
In Sec.~\ref{sec:insp_only} we compare the inspiral portion of EOBNR waveforms to HG and TaylorT4 
waveforms. In Sec~\ref{sec:conclusion} we consider the implications of our results for 
future searches in advanced detectors. 

\section{Waveforms}

In this section we summarise the key concepts entering the construction of the waveforms
used in this study. Throughout the paper, for a binary system with individual component masses $m_1$ and $m_2$ (with $m_{2} < m_{1}$) we define the total mass as $M \equiv m_{1}+m_{2}$, and  mass ratio and symmetric mass ratio as $ q \equiv m_2/m_1$ and $\eta \equiv m_{1}m_{2}/(m_{1}+m_{2})^{2}$, respectively.

We consider the family of waveforms constructed by calibrating the effective-one-body approach
to numerical relativity (EOBNR) \cite{PanEOBNR:2011}.  The EOBNR family describes the full inspiral-merger-ringdown signal; it is
currently used in searches that reach the IMBH mass range, so far up to \unit{100}{\smass} \cite{s6Highmass:2012}.
The free parameters in the family have been fitted to comparable mass ratio numerical relativity simulations, and
by construction this family is deemed to be faithful in the test particle limit.
For this work, we use the implementation provided by the LIGO Scientific Collaboration 
Algorithm Library (LAL) that corresponds to the approximant EOBNRv2.
% We note that there also exists another waveform family which also describes the full inspiral-merger-ringdown signal known as IMR Phenom-B \cite{phenB}. This is a frequency domain waveform family constructed by smoothly matching PN inspiral waveforms to numerical relativity simulations.However, unlike EOBNR it is not designed to be faithful in the extreme mass-ratio limit and hence may not be appropriate for our study. We therefore use only use EOBNR as the inspiral-merger-ringdown waveform in our study.

We also consider a waveform family based on test particle motion in Kerr/Schwarzschild space-time with 
radiative and conservative self-force corrections which we refer to as the Huerta-Gair (HG) 
family~\cite{HG:2011}. The approximation scheme is constructed specifically 
to handle highly-asymmetrical mass-ratio binaries and is therefore a physically well motivated 
approximation scheme for intermediate mass-ratio inspirals. These waveforms have been compared against, Teukolsky based
waveforms for inspiralling test particles on geodesic orbits and the match exceeds $95\%$ over a large portion of the parameter space \cite{HuertaGair:2009}.
These waveforms have been used to study the detection of intermediate mass-ratio inspirals by the Einstein Telescope in \cite{HG:2011}.
The Huerta-Gair waveforms describe only the inspiral portion of the coalescence signal.
There is no corresponding LAL approximant. The gravitational-wave polarization states can be computed from Eqs.~(14) and (15) of \cite{HG:2011}. For our study, effects of orientation of the gravitational-wave source are irrelevant and we can consider only circularly-polarized face-on binaries. We fix the spin parameter to zero.

Finally, as a reference we use the standard inspiral-only post-Newtonian approximation corresponding
to the LAL approximant TaylorT4, which includes corrections to the phase of the waveform at 3.5PN order~\cite{Buonanno:2009}.

The TaylorT4 waveforms used here are computed in the so-called ``restricted'' amplitude approximation, which assumes the waveform amplitude to be zeroth post-Newtonian order and only includes the leading second harmonic of the orbital phase. We only include the leading second harmonic of the orbital phase in the EOBNR waveforms. We do not consider the effects of spin or eccentricity in any of the waveform families, as we restrict to circular orbits and non-spinning black holes. %Spins and/or eccentricity may be non-negligible for binaries with IMBHs (need refs) and the initial work presented here will need to be extended to include these effects.

The HG and TaylorT4 families are inspiral-only time-domain waveforms and are terminated when the gravitational waveforms reach the ISCO frequency.

\label{sec:waveforms}
\section{SNR from inspiral, merger and ringdown}

In this section we consider the relative contributions of the different portions of the gravitational-wave coalescence signal to the SNR as a function of the IMRAC's mass. 

We work in the frequency domain and define the Fourier transform of the gravitational-wave strain signal,  $\tilde{h}(f)$, as
\beq\label{ft} 
\tilde{h}(f) = \int^{+\infty}_{-\infty} dt\ h(t)e^{-2\pi i ft}\,,
\eeq
where $h(t)$ is the time-domain strain signal. We define the noise-weighted inner product as
\beq
(a|b) = 4\Re\left[\int^{f_{\mathrm{max}}}_{f_{\mathrm{min}}}df \ \frac{\tilde{a}(f)\tilde{b}^{*}(f)}{S_{n}(f)}\right],\
\eeq
\noindent where $S_{n}(f)$ is the instrument noise power spectral density (PSD), which we will take to be the 
Advanced LIGO high-power, zero-detuned noise PSD \cite{aLIGOpsd}. The limits of integration correspond to the bandwidth
of the detector. The expectation value of the optimal matched filtering SNR, in the case when the signal and template waveforms are identical, is given by \cite{CutFlan:93}
\bea
\Big(\frac{S}{N}\Big)_{\mathrm{max}} & = & (h|h)^{1/2}\,,
\nonumber\\
& = & \Bigg[4\Re\int \left(\frac{f |\tilde{h}(f)|}{\sqrt{f S_n(f)}}\right)^2 d\ln f\Bigg]^{1/2}.\ 
\label{eq:max_snr} 
\eea
Writing the maximum SNR in the form above clearly separates it into contributions from the signal strain, $f |\tilde{h}(f)|$, and the root-mean-squared (rms) noise spectral amplitude, $\sqrt{f S_n(f)}$, which is the strain signal associated with the detector noise.

\begin{figure*} 
\includegraphics[scale=0.5]{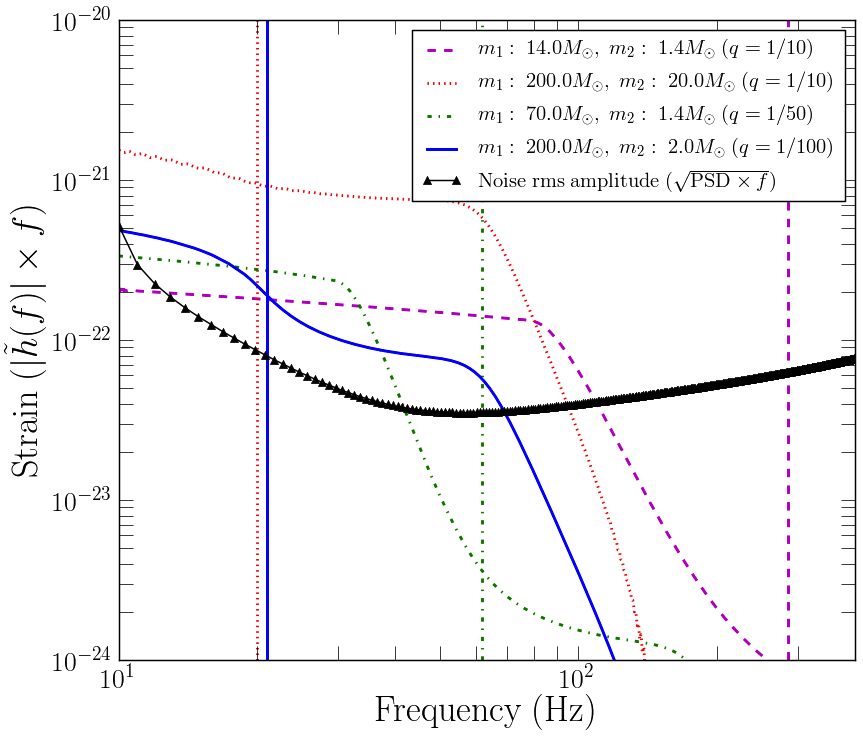} 
\caption{Strain of optimally-located and oriented IMRAC sources at a fiducial distance of 1 Gpc as described by EOBNR waveforms, and aLIGO noise. The corresponding ISCO frequency 
of each signal is shown as a vertical line. The strain from the merger and ringdown from each source contributes after the ISCO frequency. For the sources with 
component masses $(m_{1}, m_{2}) = [\unit{(200, 20)}{\smass}, \unit{(200, 2)}{\smass}]$, the strain from 
merger and ringdown sits above the noise spectrum. The SNR from the full EOBNR waveform and from its inspiral-only portion are shown in Fig.~\ref{fig:SNRs}.}
\label{fig:RMS} 
\end{figure*}

One can gain insight into the relative contributions to the SNR  
from inspiral, merger and ringdown by comparing the gravitational-wave strain to the noise rms value.
In Fig.~\ref{fig:RMS} we show the strain for selected overhead and face-on (i.e., optimally-located and oriented) 
IMRAC sources at a fiducial distance of 1Gpc as described by the EOBNR waveform family, and noise rms amplitude. 
The ISCO frequency of each signal is shown as a vertical line. The strain from 
merger and ringdown is thus the portion after the ISCO frequency. The contribution to the 
strain from merger and ringdown from binaries with component masses $(m_{1}, m_{2}) = [\unit{(200, 20)}{\smass}, \unit{(200, 2)}{\smass}]$, is greater than that of the noise rms amplitude (black curve with triangles). 
In general, systems with ISCO frequencies between \unit{30-100}{\hertz}, merge in the ``bucket'' of the noise curve, i.e., where the detector is 
most sensitive. For example, for the $(m_{1}, m_{2}) 
= \unit{(200, 20)}{\smass}$ system (red dotted curve in Fig.~\ref{fig:RMS}), the merger and ringdown contribute the 
bulk of the SNR. Conversely, the inspiral contribution to the SNR is strongly suppressed for such massive systems.

\begin{figure*} 
	\centering 
	\begin{subfigure}
			\centering 	
			\includegraphics[scale=0.14]{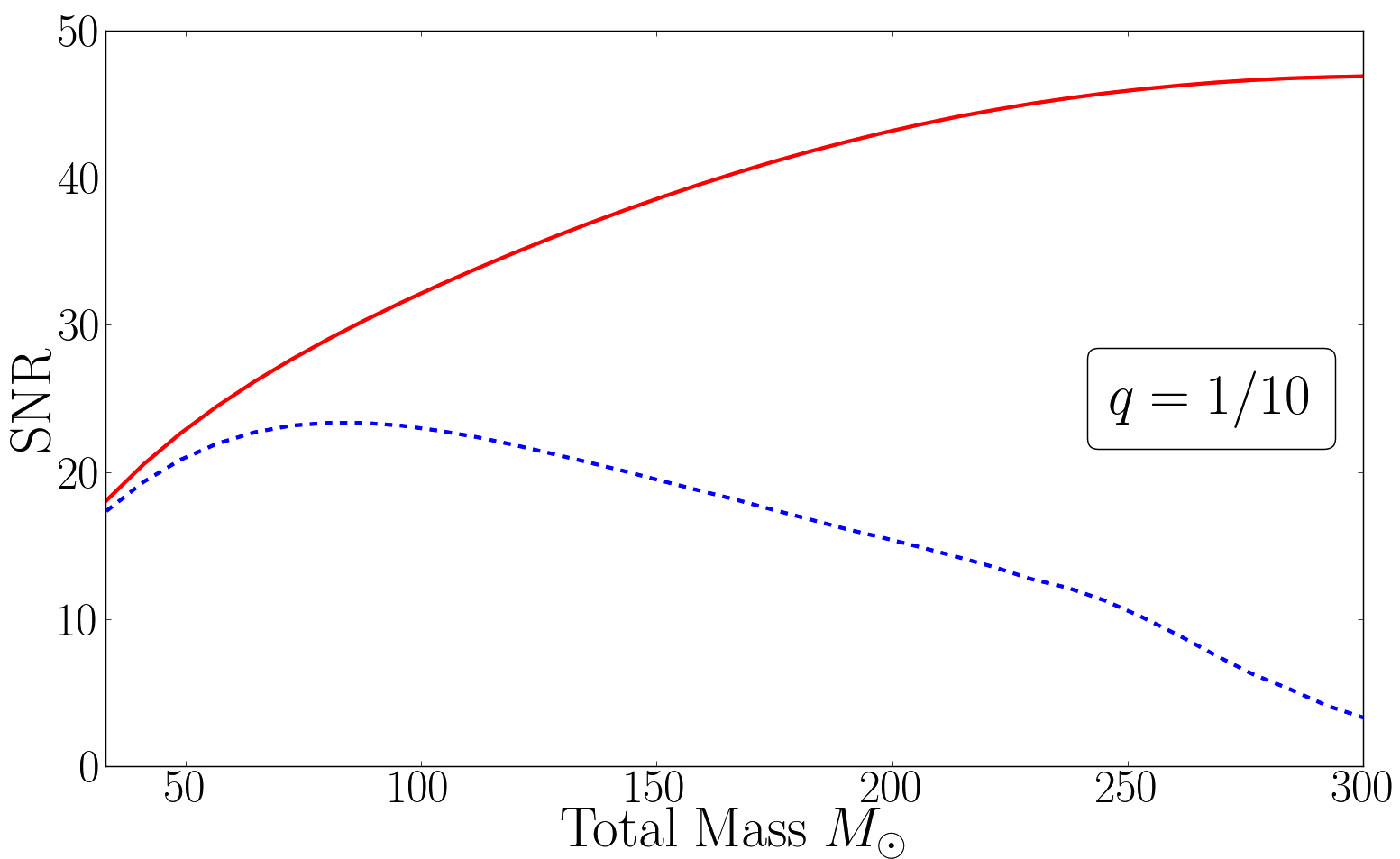}
	\end{subfigure} 
	\begin{subfigure} 
			\centering 
			\includegraphics[scale=0.14]{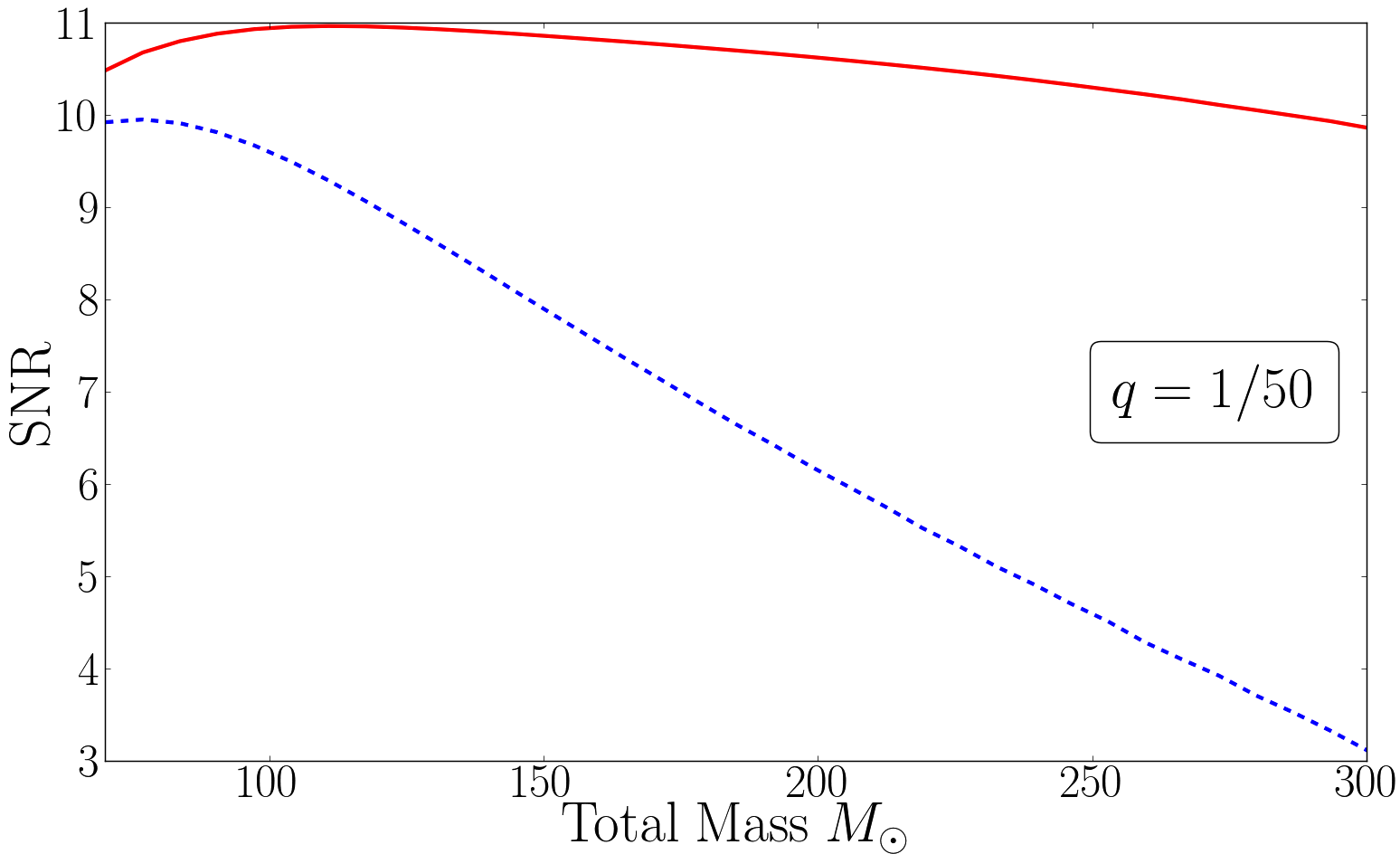} \\
	\end{subfigure} 
	\begin{subfigure} 	
			\centering 
			\includegraphics[scale=0.14]{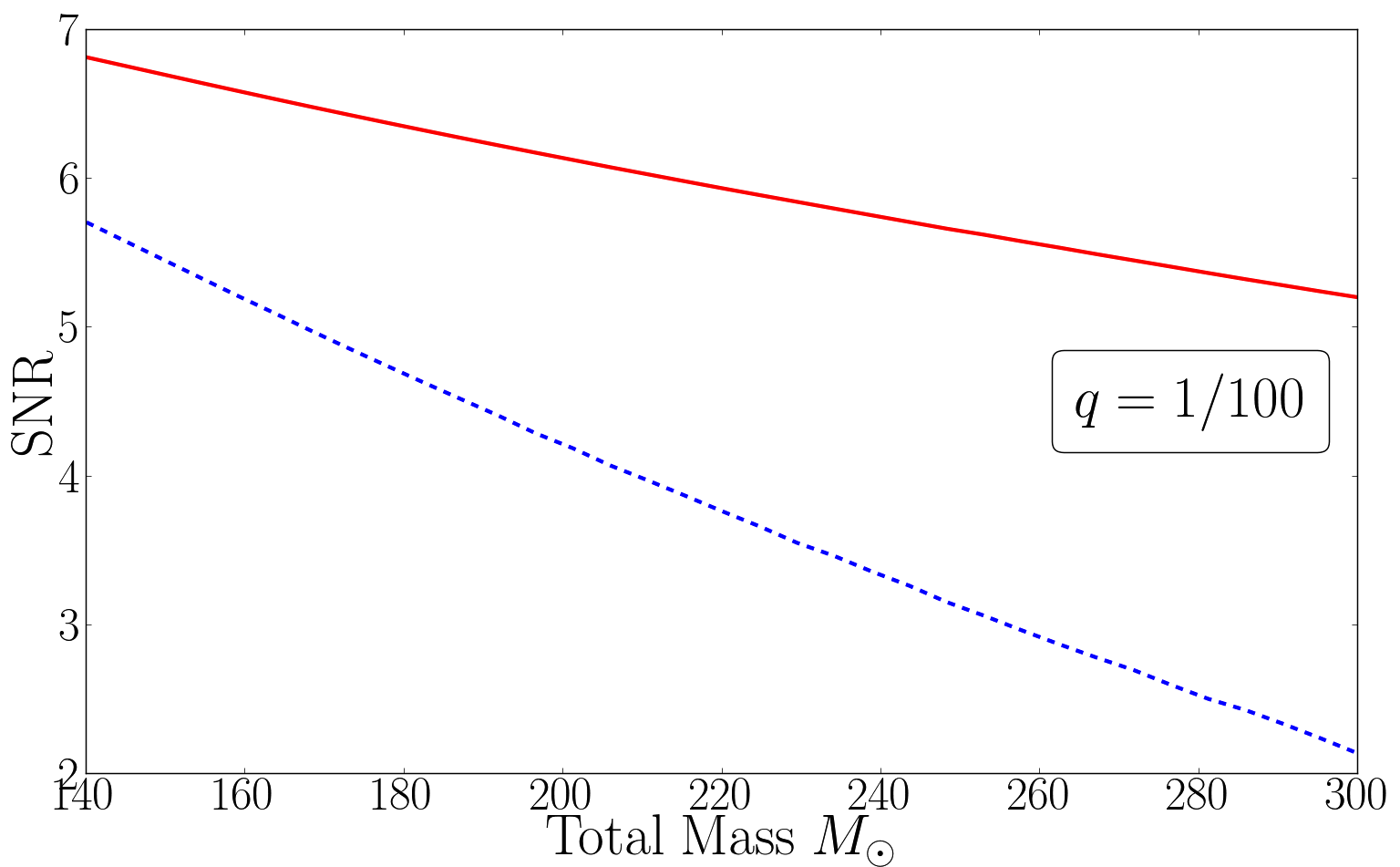} 
	\end{subfigure} 
	\begin{subfigure}
			\centering 
			\includegraphics[scale=0.14]{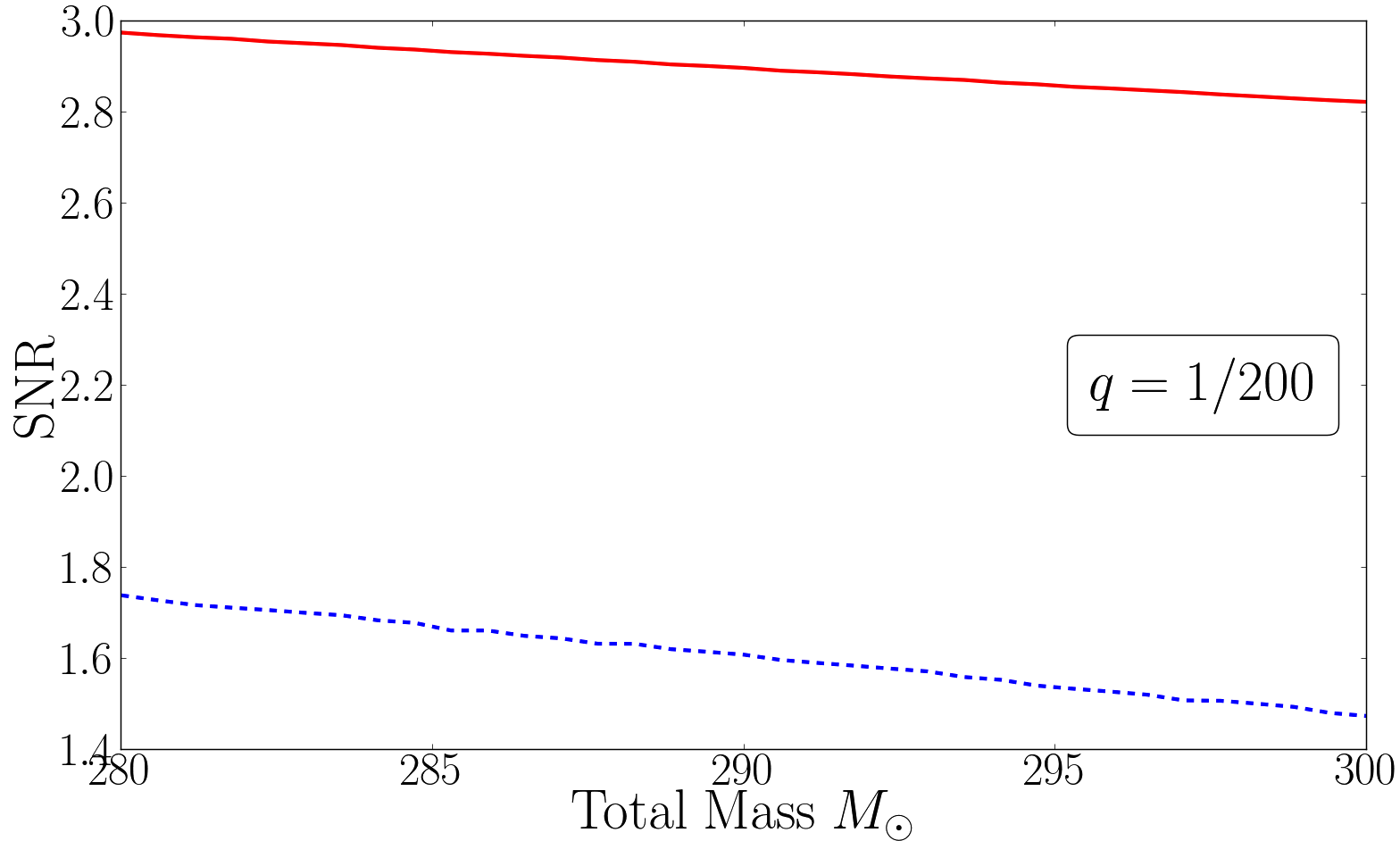} 
	\end{subfigure} 
\caption{SNR of optimally-located and oriented IMRAC sources at a fiducial distance of 1 Gpc vs total mass for four different mass ratios; $q= 1/10, 1/50, 1/100, 
1/200$. The solid line 
is the SNR from full EOBNR waveforms and the dashed line from inspiral-only EOBNR waveforms truncated at the ISCO frequency in the frequency domain. The lower-bound mass of the smaller body is set to $m_{2} = \unit{1.4}{\smass}$ which is the canonical neutron-star mass. The lowest total mass for the $q=1/50,\ 1/100$ and $1/200$ subplots in Fig.~\ref{fig:SNRs} is set by fixing the mass of the smaller body to $m_{2} = \unit{1.4}{\smass}$. For the $q=1/10$ subplot the smallest total mass is set to $M=$ \unit{35}{\smass} as the inspiral accounts for the vast majority of the SNR below this value.  We only consider systems with total masses such that the ISCO 
frequency is greater than \unit{10}{\hertz} (our low frequency cut-off). The highest total mass for each of the subplots in Fig~\ref{fig:SNRs} is set to $M=$ \unit{300}{\smass} which ensures that the ISCO 
frequency is greater than \unit{10}{\hertz}. We find that there is a non-negligible contribution to the SNR from             
merger and ringdown in IMRAC signals above a total mass of around $M=$ \unit{35}{\smass}. The difference
in SNR between inspiral-only and full waveforms is at the $3\%$ level at around $M=$ \unit{35}{\smass} at $q=1/10$.
For binaries with $q=1/50, 1/100\ \mathrm{and}\ 1/200$, the minimum loss in SNR are at the $6 \%$, $15\%$ and $40\%$ levels, respectively, in our mass range of interest.
For IMRACs of astrophysical interest, more extreme mass ratios correspond to greater total mass, which can place the merger and ringdown at a frequency where the detector has the greatest sensitivity.} 
\label{fig:SNRs} 
\end{figure*}

In Fig.~\ref{fig:SNRs} we show the maximum SNR, Eq.~(\ref{eq:max_snr}), as a function of the binary's total mass produced by
inspiral-only and full EOBNR waveforms at four different mass-ratios in the range $1/200 \le q \le 1/10$. 
We construct inspiral-only EOBNR waveforms by Fourier transforming the full waveform into the 
frequency domain and truncating it at the ISCO frequency.
We have considered the SNR for optimally-located and oriented sources at a fiducial distance of 1 Gpc. The lower-bound mass of the smaller body is set to $m_{2} = \unit{1.4}{\smass}$ which is the
canonical neutron-star mass. The lowest total mass for the $q=1/50,\ 1/100$ and $1/200$ subplots is set by fixing the mass of the smaller body to $m_{2} = \unit{1.4}{\smass}$. For the $q=1/10$ subplot in Fig.~\ref{fig:SNRs} the smallest total mass is set to $M=$ \unit{35}{\smass} as the inspiral phase accounts for the vast majority of the SNR below this value.  The lower limit of 
integration of Eq.~(\ref{eq:max_snr}) is \unit{10}{\hertz} and the upper limit is \unit{2048}{\hertz}, which is the Nyquist frequency of discretely sampled EOBNR waveforms generated at a sampling rate of $\Delta\,t = 4096\,$s in the time-domain.  We only consider systems with total masses such that the ISCO 
frequency is greater than \unit{10}{\hertz} (our low frequency cut-off). The highest total mass for each of the subplots in Fig~\ref{fig:SNRs} is set to $M=$ \unit{300}{\smass} which ensures the ISCO 
frequency is greater than \unit{10}{\hertz}.

As anticipated from Fig.~\ref{fig:RMS} there is a significant difference
in the SNR between inspiral-only and full EOBNR waveforms that can be seen at all four mass ratios. 
We also note that for systems with mass ratios of $q=1/10$ with total masses below 
around $M=$ \unit{35}{\smass} the inspiral phase is the dominant source of SNR. If we consider 3\%
as a fiducial value of the difference between the full SNR and the one associated to the inspiral-only waveform -- which leads to a loss in detection rates of $10\%$ -- this happens at 
$M\approx$ \unit{35}{\smass} for $q=1/10$.
For binaries with $q=1/50,\ 1/100$ and $1/200$, the minimum difference in SNR between inspiral-only and full waveforms is  $\approx 6\%,\ 15\%$ and $40\%$ respectively for the mass ranges considered in Fig~\ref{fig:SNRs}.

In summary, we have shown that inspiral-only templates will miss a significant portion of the total SNR of IMRAC signals
over the bulk of the detectable mass-space. Future searches will therefore require templates that can match the full inspiral-merger-ringdown. 
However, there is a small region of the parameter space for which inspiral-only templates may suffice for searches, without inducing drastic losses in detection rates. In the following section we quantify the effectiveness of inspiral-only templates for searching for full coalescence signals in aLIGO.

\label{sec:SNR}

\section{Effectiveness of inspiral-only templates for IMRAC searches}

We have shown in Sec.~\ref{sec:SNR} that the SNR from merger and ringdown will provide 
a significant contribution to the total SNR over a broad portion of the IMRAC mass-space, c.f. 
Fig.~\ref{fig:SNRs}. There is however a small portion of the parameter space where the SNR 
is dominated by the inspiral phase. This can be seen in Fig.~\ref{fig:SNRs} for binaries with $q=1/10$ binaries in
Fig.~\ref{fig:SNRs} with total masses $M\leq$ \unit{35}{\smass}. Thus it is important to quantify the effect of using inspiral-only 
templates to search for IMRAC signals which contain inspiral, merger and ringdown 
phases. 
The use of 
template waveforms that are not exact representations of the signals they filter degrades the SNR,
as the optimal SNR can be recovered only
when the template waveform
corresponds exactly to $h$, see ~Eq.~(\ref{eq:max_snr}). In practice however, we 
do not have access to an exact representation for $h$. Using a non-exact template 
waveform $T$ to filter $h$ caps the maximum recoverable SNR to
\bea\nonumber
\Big(\frac{S}{N}\Big) &=& \max_{\vec{\theta}} \frac{(h(\vec{\lambda}) | T(\vec{\theta}))} 
{(T(\vec{\theta})|T(\vec{\theta}))^{1/2}}\,,\\
&=&\epsilon \ \Big(\frac{S}{N}\Big)_{max}\,, 
\eea
where $\vec{\lambda}$ and $\vec{\theta}$ represent the parameter vector of the signal and template, respectively. We define $\epsilon$  as the
\textit{effectiveness} of a template waveform family $T$ at recovering the maximum SNR from a gravitational-wave signal $h$; by definition $0 \le \epsilon \le 1$. This quantity is also referred to as the ``fitting factor" in the literature \cite{Apo:1995}. It is convenient 
to define waveforms normalized to unit norm as $\mathbf{\hat{a}}(f)=\tilde{a}(f)/(a|a)^{1/2}$ so that $(\mathbf{\hat{h}}|\mathbf{\hat{h}})= 
(\mathbf{\hat{T}}|\mathbf{\hat{T}})=1$ and the effectiveness can be written succinctly as \cite{Apo:1995}
\beq 
\epsilon = \max_{\vec{\theta}} (\mathbf{\hat{h}}(\vec{\lambda})|\mathbf{\hat{T}}(\vec{\theta}))\,.
\label{epsilon} 
\eeq
Using normalized waveforms also has the advantage of eliminating the 
dependence of the waveforms on the source orientation and distance, which enter as an 
overall scaling.

Calculating the effectiveness, Eq. (\ref{epsilon}), requires maximizing over the 
component masses ($m_{1}, m_{2}$) and the time and phase at coalescence. We can 
efficiently maximize over the time and phase by Fourier transforming the integrand of the noise-weighted inner-product \cite{findchirp:2005}, 
\beq 
z(t_{c}) =  4\int^{f_{max}}_{f_{min}}df\ \frac{\tilde{a} 
(f)\tilde{b}^{*}(f)}{S_{n}(f)}\ e^{2\pi ift_{c}},\ 
\eeq
which yields a complex time series whose elements correspond to the inner-product of $a$ and $b$ as one of the signals is time-shifted with respect to the other.
We can efficiently find the time at coalescence, $t_{c}$, by finding the time at which the norm of this time series is a 
maximum. The phase at coalescence $\phi_{c}$ is then automatically given by finding the 
argument of the time-series at its peak amplitude. We thus modify the inner product 
$(a|b)$:
\beq 
(a|b) \rightarrow  (a|b)^{\prime} = \max_{t_{c}}\left|z(t_{c})\right|,\ 
\label{eq:mod_innerprod} 
\eeq
which we will adopt as the definition of the inner-product for the remainder of this paper. 

\begin{figure*} 
\includegraphics[scale=0.3]{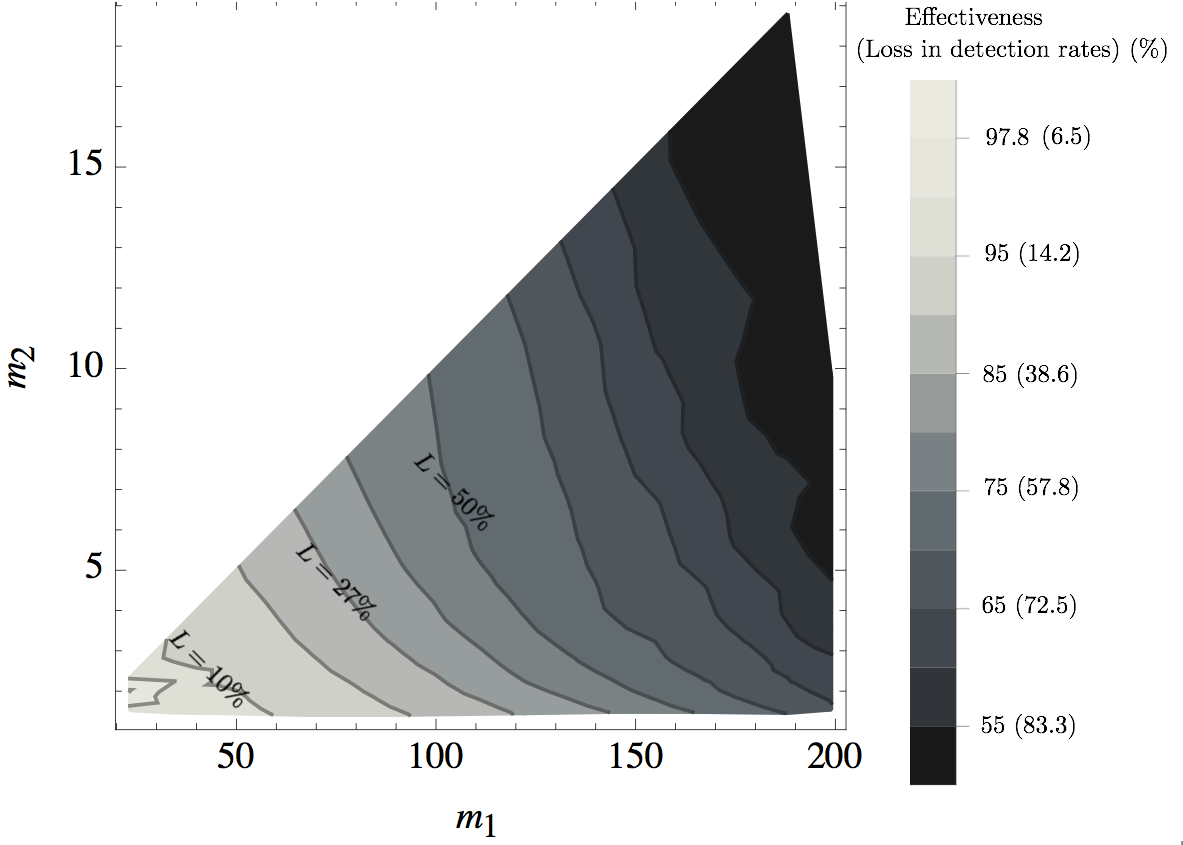}
\caption{Effectiveness of inspiral-only EOBNR templates to filter full inspiral, merger and 
ringdown EOBNR signals as a function of the source component masses and 
corresponding loss in detection rates. The diagonal corresponds to a mass ratio 
$q=1/10$. Inspiral-only EOBNR templates are constructed by truncating the full waveform 
at the ISCO frequency in the frequency-domain. For the bulk of the parameter space 
inspiral-only templates are $\lesssim 75\%$ effective at filtering inspiral, merger and ringdown signals. Inspiral-only templates are $\sim 97-98\%$ effective for total masses 
$M_{\odot} \lesssim$ \unit{30}{\smass}. Inspiral-only templates within the $90\%$-effectiveness contour should be sufficient for IMRAC searches without incurring greater than $30\%$ losses in detection rates. The contours corresponding to losses in detection rates $L=10\%$, $L=27\%$ and $L=50\%$ have been highlighted.
}
\label{fig:ff_eobnrs_rates} 
\end{figure*}

To compute the effectiveness of an inspiral-only IMRAC search we evaluate Eq.~(\ref{epsilon}) 
for signals covering the IMRAC mass space. We take as our signal 
waveform, $h$, the full inspiral-merger-ringdown EOBNR waveform. We take the template, $T$, to be an \textit{inspiral-only}
EOBNR waveform, formed by truncating the full EOBNR waveform at the ISCO frequency in the frequency domain. 
With such signals and templates the effectiveness provides a measure of the maximum SNR 
which could be achieved through using an inspiral-only template to filter full coalescence-signals. 
To get a broad coverage of the IMRAC mass space we compute Eq.~(\ref{epsilon}) for 
signals whose source masses cover the ranges \unit{1.4}{\smass} $\leq m_{2} \leq$ 
\unit{18.5}{\smass} and \unit{24}{\smass} $\leq m_{1} \leq$ \unit{200}{\smass}, with mass ratios spanning the range $q:=m_{2}/m_{1} 
\in [1/140, 1/10]$. For each signal we evaluate the effectiveness, Eq.~(\ref{epsilon}), where the template $T$ describes the inspiral-only portion of an EOBNR waveform. 
We maximize over time and phase by maximizing the inner product of the signal with an inspiral-only EOBNR template, Eq.~(\ref{eq:mod_innerprod}). 
The maximization over the masses is performed by finding the largest inner product between the 
signal and a bank of template waveforms. The template bank is characterised by intrinsic parameters $(\mathcal{M}\,,\eta)$ which are the chirp mass and symmetric mass ratio of the system: $\mathcal{M} = (m_1m_2)^{3/5}/(m_1 + m_2)^{1/5}$, $\eta = (m_1m_2)/(m_1 + m_2)^{2}$, where $m_1$ and $m_2$ are the component masses of the binary. The bank spans an extended mass range $3\,M_{\odot} \leq \mathcal{M}\,\leq30\,M_{\odot}$ and $ 0.0065\,\leq\, \eta\,\leq 0.082$.
The results of the effectiveness of an inspiral-only IMRAC search are shown in Fig.~\ref{fig:ff_eobnrs_rates}.

Inspiral-only templates are $\sim 98\%$ effective at filtering full coalescence signals 
for total masses $M \lesssim$ \unit{30}{\smass}. Such systems have an ISCO frequency 
\unit{150}{\hertz} $\lesssim f_{\mathrm{ISCO}}$ which is well within the peak sensitivity of the noise 
curve. However for the bulk of the mass space the effectiveness is 
below $75\%$. This is unsurprising given the SNR curves in Fig.~\ref{fig:SNRs} which 
clearly show the importance of the contribution of merger and ringdown to the SNR.

The loss in SNR incurred through using inspiral-only templates directly affects 
detection rates. Because the SNR scales inversely with the distance, the observable volume will scale with the cube of the effectiveness. Assuming that GW sources are isotropically distributed in the sky, the 
fractional loss in detection rates will be $L = 1-\epsilon^{3}$, assuming that detectability is determined by SNR threshold alone.  The percentage loss in detection rates through using 
inspiral-only EOBNR templates to recover the full coalescence signal is also shown in Fig.~\ref{fig:ff_eobnrs_rates}. Over a broad portion of 
the mass-space inspiral-only templates incur losses in detection rates between $60- 
85\%$. As the total mass of the binary approaches \unit{440}{\smass} the Schwarzschild ISCO frequency, Eq.~(\ref{eq:fisco}), approaches \unit{10}{\hertz} which is near the low-frequency cut-off of the detectors. Hence the relative contribution of the inspiral phase to the coalescence signal of heavier systems diminishes until the only contribution is from the merger and ringdown. This is a striking indication of the need of merger and ringdown in IMRAC 
template waveforms. This suggests the importance of full numerical simulations in this 
regime in order to construct a reliable waveform family including inspiral, merger, and ringdown phases.

\label{sec:IMR}

We identify three regions in the $m_{1}$-$m_{2}$ plane in which various searches 
could be constructed. The regions are defined by contours of constant effectiveness which are approximately
given by $\mathcal{C} = (m_{1}/M_{\odot})\sqrt{m_{2}/M_{\odot}}$, which are found purely empirically, with $\unit{1.4}{\smass} \leq m_{2} \leq 
\unit{18.5}{\smass}$ and mass ratios $q \in [1/140, 1/10]$. The effectiveness is related to $\mathcal{C}$ by 
$\epsilon \approx 1/100\times (1.6\ \mathcal{C} - 7.3\times10^{-3}\ \mathcal{C}^{2})$.

Between the $97\%$- and $90\%$-effectiveness contours, the losses in 
detection rates are between $10\% \lesssim L \lesssim 27\%$ and so an inspiral-only search could be sufficient without 
incurring drastic losses in detections. The region bound from below in effectiveness by the $90\%$-effectiveness contour is defined by 
$\mathcal{C} \leq 100$, with the effectiveness increasing with decreasing $\mathcal{C}$. Between the $90\%$- and $80\%$- 
effectiveness contours the losses in detection rates are around $27\% \lesssim L \lesssim 50\%$. Thus, within 
this region searches will be limited by the lack of merger and ringdown in template 
waveforms, though an inspiral-only search would be feasible in principle. This contour is defined by $100 \lesssim \mathcal{C} \lesssim 150$. 
Below the $80\%$-effectiveness contour, inspiral-only searches will incur losses in 
detection $50\% < L$ and so merger and ringdown will be crucial for searches. The region 
bound from above in effectiveness by the $80\%$-effectiveness contour which is defined by $150 \lesssim \mathcal{C}$, with effectiveness decreasing with increasing $\mathcal{C}$. The results 
are summarized in Table~\ref{table:search_regions}. We note that these contours are accurate to within around $5\%$ of the true value of the constant-effectiveness contours between the $L=10\%$ and $L=50\%$ contours. Below the $L=10\%$ and beyond the $L=50\%$ contours the accuracy deteriorates, however we do not consider individual contours outside these ranges.

\begin{table*}[!htp] 
\begin{center} 
\begin{tabular}{ | p{3cm} | p{3.cm} | p{4.8cm} |p{5.cm} |} 
\hline 
Effectiveness of inspiral-only search, $\epsilon(\%)$  & Loss in detection rates, $L(\%)$& 
Contours of constant effectiveness in $m_{1}$-$m_{2}$ plane $(\mathcal{C} = (m_{1}/M_{\odot})\sqrt{m_{2}/M_{\odot}})$ within mass range of interest & Implication for searches\\ \hline 
$ 90\% \lesssim \epsilon  \lesssim 97\%$ & $ 10\% \lesssim L  \lesssim 27\%$ & 
$\mathcal{C} \lesssim 100$ & Inspiral-only search sufficient but 
with non-negligible loss in detection rates.\\ \hline 
$80\% \lesssim \epsilon  \lesssim 90\%$ & $27\%  \lesssim L  \lesssim 50\%$ & $100  \lesssim
\mathcal{C} \lesssim 150$ & Inspiral-only search possible but 
limited by lack of merger and ringdown. Could potentially lose half of signals with 
inspiral-only templates.\\ \hline 
$\epsilon \lesssim 80\%$ & $50\% \lesssim L $ & $150 \lesssim \mathcal{C}$ 
& Merger and ringdown crucial for searches. Will miss over half of signals with 
inspiral-only templates.\\ \hline 
\end{tabular} 
\caption{Effectiveness of inspiral-only searches, the corresponding loss in detection 
rates and the region in the $m_{1}$-$m_{2}$ plane bounded by constant-effectiveness 
contours. For a given region in the $m_{1}$-$m_{2}$ plane bounded by constant-effectiveness contours we summarize the implications for IMRAC searches.} 
\label{table:search_regions} 
\end{center} 
\end{table*}

\subsection{Bias in parameter estimates from inspiral-only waveforms}
\label{sec:param_bias}
The parameters of the binary are encoded in the gravitational-wave signal.  The template gravitational waveform which maximises the SNR yields the best-fit estimate of the binary's parameters.  Neglecting the merger and ringdown in template waveforms can lead to biases in the best-fit parameters when the signal describes the full inspiral, merger and ringdown phases of coalescence.  Here, we analyze biases in the mass parameters: the chirp mass $\mathcal{M}$ and the symmetric mass ratio $\eta$.

\begin{figure*} 
\includegraphics[scale=0.34]{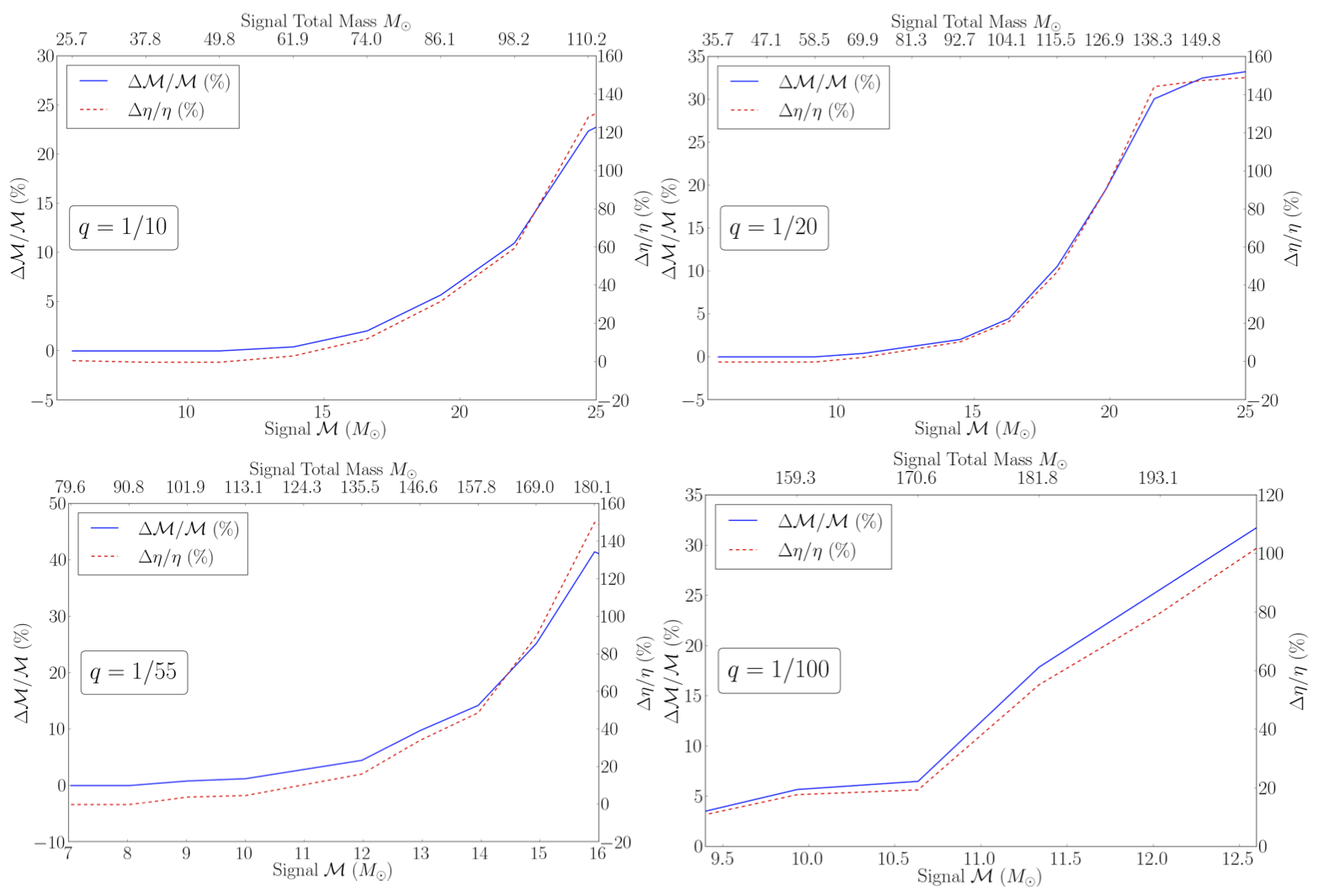} 
\caption{
Biases in best-fit estimates, incurred through using inspiral-only template waveforms, of chirp mass $\mathcal{M}$ and symmetric mass ratio $\eta$ as a function of chirp mass of a signal waveform that describes the inspiral, merger and ringdown phases of coalescence. Biases are shown for four different mass ratios of the signal waveform, $q=1/10\,, 1/20\,, 1/55\,, 1/100$. The biases $\Delta\mathcal{M}/\mathcal{M}$ and $\Delta\eta/\eta$ are measured as the ratio of the difference between the best-fit and the signal parameter to the signal parameter: $ \Delta\mathcal{M}/\mathcal{M} = (\mathcal{M}^{\mathrm{best\ fit}} - \mathcal{M})/\mathcal{M}$ and $\Delta\eta/\eta = (\eta^{\mathrm{best\ fit}} - \eta)/\eta$. For convenience we have also shown the total mass of the signal on the upper x-axis.
}
\label{fig:bias} 
\end{figure*}

In Fig.~\ref{fig:bias} we compare biases in the best-fit values of chirp mass and symmetric mass ratio obtained with inspiral-only template waveforms as a function of the chirp mass encoded in full EOBNR signal waveforms. We report the fractional bias in the chirp mass and mass ratio: $ \Delta\mathcal{M}/\mathcal{M}= (\mathcal{M}^{\mathrm{best\ fit}} - \mathcal{M}^{\mathrm{true}})/\mathcal{M}^{\mathrm{true}}$ and $\Delta\eta/\eta = (\eta^{\mathrm{best\ fit}} - \eta^{\mathrm{true}})/\eta^{\mathrm{true}}$.

These systematic biases should be compared against typical statistical measurement uncertainties.  However, approximate techniques for estimating measurement uncertainty from the Fisher information matrix are notoriously unreliable \cite{Vallisneri:Fisher1, Vallisneri:Fisher2}, and a full Bayesian treatment of IMRAC parameter estimation is beyond the scope of this paper.  We note, however, that for comparable-mass binaries, the chirp mass and symmetric mass ratio can typically be measured to within $\mathcal{O}(1\%)$ and $\mathcal{O}(10\%)$ of the signal value \cite{S6PE}.

For the $q=1/10$ and $q=1/20$ cases we find that biases in the best-fit parameters can be as low as $\Delta\mathcal{M}/\mathcal{M}\, \sim\,0.1\%$ and $0.1\% \lesssim \Delta\eta/\eta \lesssim 10\%$ for values of the signal chirp mass $\mathcal{M}\lesssim 15\,M_{\odot}$. These biases are lower than typical measurement uncertainty. However, the biases in chirp mass and symmetric mass ratio rise significantly for $\mathcal{M} \gtrsim 15\, M_{\odot}$ for the $q=1/10$ and $q=1/20$ cases: the biases in chirp mass can be as high as $20\%$ and $30\%$ respectively, and those for symmetric mass ratio as high as $120\%$ and $160\%$ respectively. Such high errors are due to the increased importance of merger and ringdown in the signal waveform at higher chirp mass.

For the $q=1/55$ case we find similar behaviour in the bias in best-fit chirp mass and symmetric mass ratio, although the smallest biases in symmetric mass ratio are of the order of $\Delta\eta/\eta \sim\,10\%$ for masses in the range of interest. In addition, we find that the bias in chirp mass is on the order of $\Delta\mathcal{M}/\mathcal{M}\, \sim\,0.1\%$ for chirp mass signal values in the range $7\,M_{\odot}\lesssim \mathcal{M}\,\lesssim 10\,M_{\odot}$.  However, the biases grow rapidly to $40\%$ when the signal value of chirp mass exceeds $10\,M_{\odot}$. For the $q=1/100$ case we find that the smallest bias in chirp mass and symmetric mass ratio are around $4\%$ and $20\%$ respectively.

%\ilya{[I would expect that total mass, or, perhaps even better, our variable $\mathcal{C}$, may be a more universal constraint on the faithfulness of inspiral-only waveforms than chirp mass alone, where the chirp mass cut clearly varies with mass ratio.  Can you please check this?]}

These results suggest that inspiral-only waveforms may only be able to provide faithful estimates of the binary's masses in some limited cases. For high-mass systems, the best-fit mass parameter estimates are likely to be highly unreliable. 

\section{Comparison of inspiral-only waveforms} 
\label{sec:insp_only}

We have shown that merger and ringdown are crucial for effective searches over a large 
portion of the IMRAC mass space, though there is a small region in which an inspiral-only 
search could be constructed without incurring losses in detection rates greater than around $27\%$. For this region, it is therefore important to study whether currently available waveforms are sufficiently accurate. The inspiral phase can be computed using perturbative expansions and thus it is 
interesting to quantify the consistency of different expansions.

To assess the effectiveness of the EOBNR inspiral, we employ a waveform family designed to approximate intermediate mass-ratio inspirals which we refer to as ``Huerta-Gair'' (HG) waveforms~\cite{HG:2011}. HG waveforms describe only the inspiral portion of the coalescence signal.
This waveform family has no corresponding LAL approximant. 

\begin{figure*} 
\centering 
\mbox{\subfigure{\includegraphics[scale=0.78]{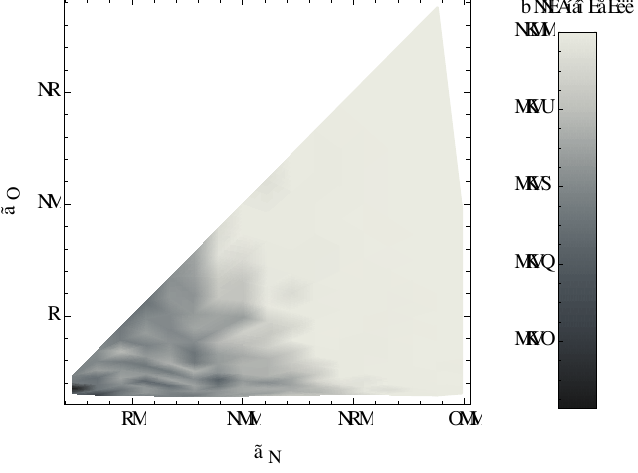}}\quad 
\subfigure{\includegraphics[scale=0.78]{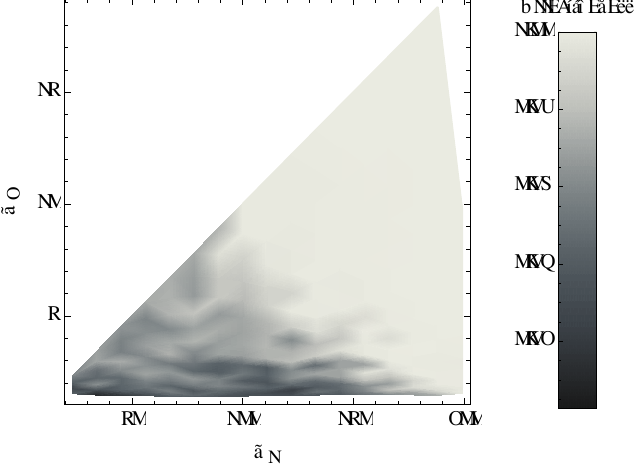} }} 
\caption{Effectiveness of inspiral-only EOBNR templates at filtering HG signals (left) and 
TaylorT4 signals (right) as a function of the source component masses encoded in the signal.} 
\label{fig:hg_tt4_eobnr} 
\end{figure*}

We repeat the study done in the previous Section using now the HG waveform family as the signal $h$ and inspiral-only EOBNR as the template $T$. The results are reported over the whole parameter space in Fig.~\ref{fig:hg_tt4_eobnr}. For completeness, in Table~\ref{table:summary} we also show the values of the effectiveness, Eq.~\ref{epsilon}, for selected mass combinations of EOBNR inspiral-only templates for filtering full EOBNR and HG signals respectively.

\begin{table}[!htp] 
\begin{center} 
%\begin{tabular}{ | p{1cm} | p{1cm} | p{2.cm} | p{2cm} |p{2.cm} |} 
\begin{tabular}{|c | c | c  c  c |} 
\hline 
$m_1$		&	$m_2$		&	\multicolumn{3}{c|}{Signal waveforms} \\
(${\smass}$)	&	 (${\smass}$)	&	full EOBNR	&	Huerta-Gair	&	TaylorT4	\\
%& & \multicolumn{3}{c|}{Signal waveforms} \\
%$\unit{m_{1}}{\smass}$ & $\unit{m_{2}}{\smass}$ & full-EOBNR signal; inspiral-only-EOBNR template & HG 
%signal; inspiral-only-EOBNR template& TaylorT4 signal; inspiral-only-EOBNR template\\ 
\hline 
50 & 5 & 0.90 & 0.95 & 0.96\\ %\hline 
100 & 5 & 0.76 & 0.97 & 0.97\\ %\hline 
200 & 5 & 0.53 & 0.99 & 0.99\\ %\hline 
50 & 1.4 & 0.96 & 0.96 & 0.90\\ %\hline 
100 & 1.4 & 0.86 & 0.98 & 0.94\\ %\hline 
200 & 1.4 & 0.67 & 0.98 & 0.98\\ %\hline 
%5 & 5 & 0.99 & 0.89 & 0.99\\ %\hline 
%20 & 20 & 0.99 & 0.92 & 0.99\\ %\hline 
%100 & 100 & 0.52 & 0.51 & 0.50\\ %\hline 
\hline 
\end{tabular} 
\caption{Summary of effectiveness of \textit{inspiral-only} EOBNR template waveforms in recovering
signals modelled using different waveform families -- full EOBNR, Huerta-Gair and TaylorT4 -- 
%with respect to various signal 
%waveforms 
for selected component masses. Merger and ringdown become more prominent in the 
coalescence signal as the total-mass of the system is increased. The EOBNR 
inspiral is typically better at matching HG signals in the IMRAC regime than TaylorT4 
signals.} %Results for equal mass-ratio systems are shown for reference below the horizontal line. Note that the effectiveness of inspiral-only EOBNR waveforms to filter full EOBNR signals when} 
\label{table:summary} 
\end{center} 
\end{table}

For the high-mass part of the mass-space the effectiveness of the EOBNR 
inspirals with respect to the HG waveforms is close to $100\%$. This is perhaps unsurprising 
because very high mass systems will have short inspirals and possible differences in the
waveforms will not produce a significant degradation of SNR when matched over a small number
of wave cycles. However, for lighter systems the effectiveness can be as low as $90\,\%$, which occurs in
the region of mass space in which inspiral-only searches would be most feasible (see Table~\ref{table:search_regions}).

%On the other hand, for the lightest systems considered here, $(m_{1}, m_{2}) = \unit{(24, 1.4)}{\smass}$, the 
%effectiveness of inspiral-only EOBNR waveforms is only $90\%$. The EOBNR waveform family has been matched to numerical 
%simulations up to mass ratios of $q = 1/6$. Meanwhile, the mass ratios considered here are not as extreme as expected for HG waveforms.  Therefore, the accuracy of both waveform families is potentially questionable in this regime.  However, the the measured effectiveness of recovering HG signals with EOBNR inspiral-only templates can be viewed as a (possibly conservative) estimate of the effectiveness of inspiral-only EOBNR waveforms at filtering the inspiral portion of whatever is nature's `true' waveform.

For reference we also compare inspiral-only EOBNR templates to TaylorT4 \cite{HG:2011} signal waveforms 
(which are inspiral-only). We construct signal waveforms on the same grid in $m_{1}- 
m_{2}$ as for HG waveforms and use the same template bank of inspiral-only EOBNR 
waveforms. The results are summarized in the right panel of Fig.~\ref{fig:hg_tt4_eobnr}, and in 
Table~\ref{table:summary} for selected masses.
We find that the EOBNR inspiral has good filtering efficiency for the TaylorT4 waveform 
family. However, EOBNR is clearly a better match to the HG waveform family over a larger range of masses and mass ratios than to TaylorT4. This can be seen more clearly by comparing the subplots in Fig.~\ref{fig:hg_tt4_eobnr}. This is unsurprising given that the PN expansion is unreliable at high velocities and highly asymmetrical mass ratios. For orbital velocities $v/c = (M\pi f)^{1/3} \gtrsim 0.2$ the PN energy flux deviates significantly from numerical results, see \cite[e.g.,][]{Poisson:1995, Yunes:2008}. A binary at its ISCO frequency has $v/c \sim 0.4$, which is well beyond the region of validity.

\section{Discussion and Conclusion} 

We have shown that over the bulk of the IMRAC mass space, merger and ringdown contribute significantly to the gravitational-wave coalescence signal.  This happens despite the suppression of the power in the merger and ringdown in signals from binaries with very asymmetric mass ratios.  The importance of merger and ringdown is due to the greater sensitivity to these waveform portions for high-mass signals, for which most of the inspiral may fall at frequencies below the detector's sensitive band.

However there is a relatively large patch in mass space in which the inspiral-only waveforms are more than $90\%$ effective. We identified three regions in which different searches 
could be considered appropriate based on thresholds of acceptable losses in detection rates. The mass space splits into a region in which inspiral-only searches
could be feasible, incurring losses in detection rates of up to $\sim 27\%$; a region in which searches would be limited by lack of merger and ringdown in template
waveforms, incurring losses in detection rates up to $50\%$; and a region in which merger and ringdown are critical to prevent losses in detection rates over $50\%$.
The search regions are summarized in Table~\ref{table:search_regions}.

We have further shown that in the region of the IMRAC mass space in which inspiral-only searches are feasible, approximants adapted to asymmetric mass-ratio
binaries are important, as here the binary is liable to have highly relativistic velocities $v/c \gtrsim 0.2$. We considered a waveform family designed to describe intermediate mass-ratio binaries which we referred to as the ``Huerta-Gair'' (HG) waveform family. By computing the effectiveness of inspiral-only EOBNR waveforms
to filter signals described by the HG waveform family, we showed that losses in recovered SNR could be as great as $10\%$. In Table~\ref{table:summary} we summarize the effectiveness of the signal--template combinations used in the paper.

We believe that template waveforms for IMRAC searches will benefit from calibration to several numerical simulations. We note that there already exists one very short numerical waveform of a $q=1/100$ binary which we have not used in our study, and which EOBNR is not currently calibrated to \cite{RIT:2011}.

In addition, it would be interesting to extend the study performed here to the case of binaries in which the components are spinning. The presence of spin in a binary leads to modulations in the emitted gravitational waveform's phase and amplitude \cite{Apo:1995}. Using template waveforms which do not account for the presence of spins can lead to further reductions in SNR \cite{SpinTbank} and systematic biases in parameter estimation, as the template waveforms will fail to account for the correct phasing of the signal waveform contained in data.

\label{sec:conclusion}

\section{Acknowledgements}
We thank Jonathan Gair for useful discussions and help in implementing the HG waveform family and Chad Hanna for useful discussions. We would also like to thank Eliu Antonio Huerta Escudero for reading a draft of the manuscript and providing us with useful comments. Research at Perimeter Institute is supported through Industry Canada and by the Province of Ontario through the Ministry of Research $\&$ Innovation. This document has LIGO document number LIGO-P1300009.

\bibliographystyle{apsrev}
\bibliography{./biblio.bib}

\end{document}